\documentclass[a4paper,10pt,oneside]{book}
\usepackage{amsmath,amssymb}
\usepackage{pdfpages}
\usepackage{a4wide}
\usepackage{enumitem}
\usepackage{mathpazo}
\usepackage{wrapfig}
\usepackage{tocloft}
\usepackage{xcolor}
\definecolor{darkred}{rgb}{0.3,0,0}
\definecolor{darkblue}{rgb}{0,0,0.3}
\definecolor{firebrick}{rgb}{0.5,0.125,0.125}
\definecolor{darkgreen}{rgb}{0,0.3,0}
\usepackage[colorlinks=true,linkcolor=firebrick,citecolor=darkgreen,urlcolor=darkblue]{hyperref} 

\newcommand{\addPaper}[3]{%
\addtocounter{chapter}{1}
\addcontentsline{toc}{chapter}{\protect{\thechapter} #2:~~\texorpdfstring{\textit{\color{darkblue}#3}}{#3}}
\includepdf[pages=-]{#1}
}

\begin{document}

\begin{center}
\def\figh{0.153}
\includegraphics[height=\figh\textwidth]{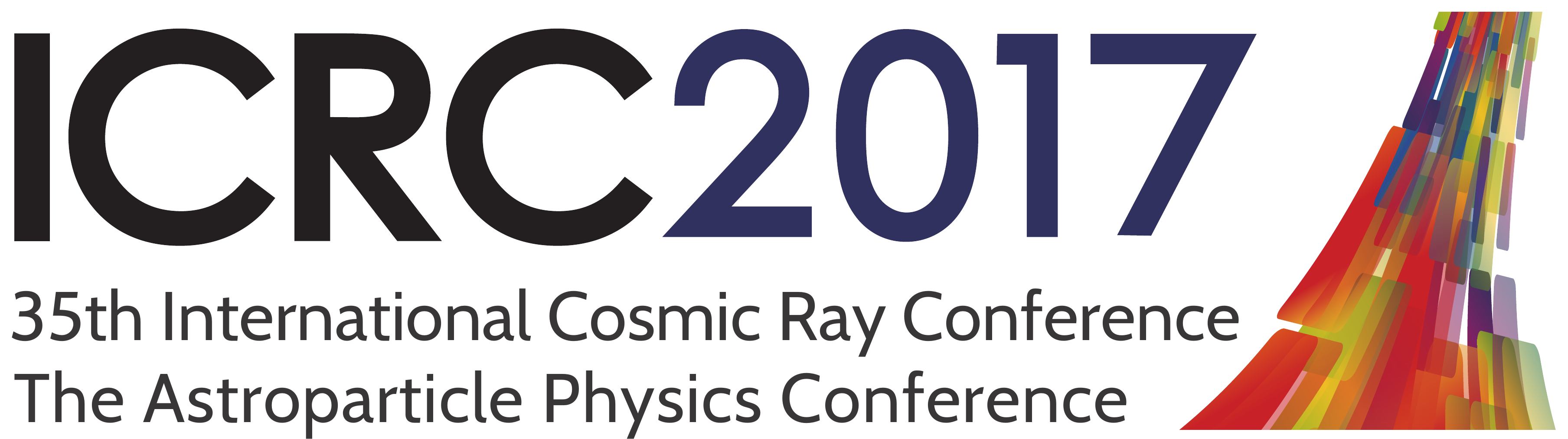}
\end{center}

\begin{center}
\Large\bf
The IceCube Neutrino Observatory, the Pierre Auger Observatory and the Telescope Array: Joint Contribution to the 35th
International Cosmic Ray Conference (ICRC 2017)
\end{center}

\begin{center}
\par\noindent
{\bf\large The IceCube Collaboration}
\end{center}

\begin{wrapfigure}[7]{l}{0.15\linewidth}
\vspace{-2.9ex}
\includegraphics[width=0.98\linewidth]{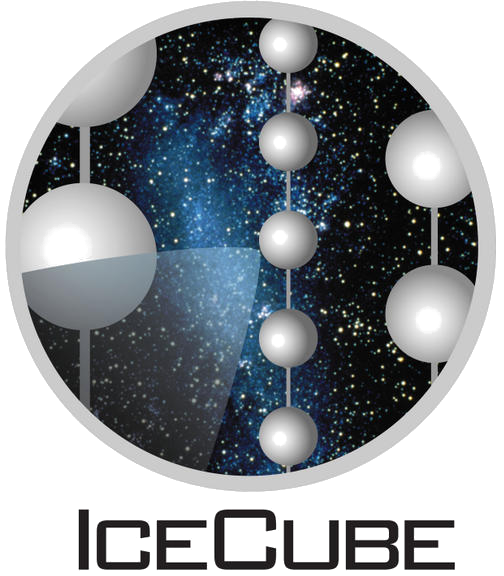}
\end{wrapfigure}
\begin{sloppypar}\noindent
M.G.~Aartsen$^{1}$,
M.~Ackermann$^{51}$,
J.~Adams$^{15}$,
J.A.~Aguilar$^{11}$,
M.~Ahlers$^{19}$,
M.~Ahrens$^{43}$,
I.~Al~Samarai$^{24}$,
D.~Altmann$^{23}$,
K.~Andeen$^{32}$,
T.~Anderson$^{48}$
% ...

\end{sloppypar}

\vspace{1ex}
\begin{center}
\rule{0.1\columnwidth}{0.5pt}
\raisebox{-0.4ex}{\scriptsize$\bullet$}
\rule{0.1\columnwidth}{0.5pt}
\end{center}

\vspace{1ex}
\begin{description}[labelsep=0.2em,align=right,labelwidth=0.7em,labelindent=0em,leftmargin=2em,noitemsep]
\item[$^{0}$] III.~Physikalisches Institut, RWTH Aachen University, D-52056 Aachen, Germany
\item[$^{1}$] Department of Physics, University of Adelaide, Adelaide, 5005, Australia
\item[$^{2}$] Dept.~of Physics and Astronomy, University of Alaska Anchorage, 3211 Providence Dr., Anchorage, AK 99508, USA
\item[$^{3}$] Dept.~of Physics, University of Texas at Arlington, 502 Yates St., Science Hall Rm 108, Box 19059, Arlington, TX 76019, USA
\item[$^{4}$] CTSPS, Clark-Atlanta University, Atlanta, GA 30314, USA
\item[$^{5}$] School of Physics and Center for Relativistic Astrophysics, Georgia Institute of Technology, Atlanta, GA 30332, USA
\item[$^{6}$] Dept.~of Physics, Southern University, Baton Rouge, LA 70813, USA
\item[$^{7}$] Dept.~of Physics, University of California, Berkeley, CA 94720, USA
\item[$^{8}$] Lawrence Berkeley National Laboratory, Berkeley, CA 94720, USA
\item[$^{9}$] Institut f\"ur Physik, Humboldt-Universit\"at zu Berlin, D-12489 Berlin, Germany
\item[$^{10}$] Fakult\"at f\"ur Physik \& Astronomie, Ruhr-Universit\"at Bochum, D-44780 Bochum, Germany
\item[$^{11}$] Universit\'e Libre de Bruxelles, Science Faculty CP230, B-1050 Brussels, Belgium
\item[$^{12}$] Vrije Universiteit Brussel (VUB), Dienst ELEM, B-1050 Brussels, Belgium
\item[$^{13}$] Dept.~of Physics, Massachusetts Institute of Technology, Cambridge, MA 02139, USA
\item[$^{14}$] Dept. of Physics and Institute for Global Prominent Research, Chiba University, Chiba 263-8522, Japan
\item[$^{15}$] Dept.~of Physics and Astronomy, University of Canterbury, Private Bag 4800, Christchurch, New Zealand
\item[$^{16}$] Dept.~of Physics, University of Maryland, College Park, MD 20742, USA
\item[$^{17}$] Dept.~of Physics and Center for Cosmology and Astro-Particle Physics, Ohio State University, Columbus, OH 43210, USA
\item[$^{18}$] Dept.~of Astronomy, Ohio State University, Columbus, OH 43210, USA
\item[$^{19}$] Niels Bohr Institute, University of Copenhagen, DK-2100 Copenhagen, Denmark
\item[$^{20}$] Dept.~of Physics, TU Dortmund University, D-44221 Dortmund, Germany
\item[$^{21}$] Dept.~of Physics and Astronomy, Michigan State University, East Lansing, MI 48824, USA
\item[$^{22}$] Dept.~of Physics, University of Alberta, Edmonton, Alberta, Canada T6G 2E1
\item[$^{23}$] Erlangen Centre for Astroparticle Physics, Friedrich-Alexander-Universit\"at Erlangen-N\"urnberg, D-91058 Erlangen, Germany
\item[$^{24}$] D\'epartement de physique nucl\'eaire et corpusculaire, Universit\'e de Gen\`eve, CH-1211 Gen\`eve, Switzerland
\item[$^{25}$] Dept.~of Physics and Astronomy, University of Gent, B-9000 Gent, Belgium
\item[$^{26}$] Dept.~of Physics and Astronomy, University of California, Irvine, CA 92697, USA
\item[$^{27}$] Dept.~of Physics and Astronomy, University of Kansas, Lawrence, KS 66045, USA
\item[$^{28}$] SNOLAB, 1039 Regional Road 24, Creighton Mine 9, Lively, ON, Canada P3Y 1N2
\item[$^{29}$] Dept.~of Astronomy, University of Wisconsin, Madison, WI 53706, USA
\item[$^{30}$] Dept.~of Physics and Wisconsin IceCube Particle Astrophysics Center, University of Wisconsin, Madison, WI 53706, USA
\item[$^{31}$] Institute of Physics, University of Mainz, Staudinger Weg 7, D-55099 Mainz, Germany
\item[$^{32}$] Department of Physics, Marquette University, Milwaukee, WI, 53201, USA
\item[$^{33}$] Universit\'e de Mons, 7000 Mons, Belgium
\item[$^{34}$] Physik-department, Technische Universit\"at M\"unchen, D-85748 Garching, Germany
\item[$^{35}$] Institut f\"ur Kernphysik, Westf\"alische Wilhelms-Universit\"at M\"unster, D-48149 M\"unster, Germany
\item[$^{36}$] Bartol Research Institute and Dept.~of Physics and Astronomy, University of Delaware, Newark, DE 19716, USA
\item[$^{37}$] Dept.~of Physics, Yale University, New Haven, CT 06520, USA
\item[$^{38}$] Dept.~of Physics, University of Oxford, 1 Keble Road, Oxford OX1 3NP, UK
\item[$^{39}$] Dept.~of Physics, Drexel University, 3141 Chestnut Street, Philadelphia, PA 19104, USA
\item[$^{40}$] Physics Department, South Dakota School of Mines and Technology, Rapid City, SD 57701, USA
\item[$^{41}$] Dept.~of Physics, University of Wisconsin, River Falls, WI 54022, USA
\item[$^{42}$] Dept.~of Physics and Astronomy, University of Rochester, Rochester, NY 14627, USA
\item[$^{43}$] Oskar Klein Centre and Dept.~of Physics, Stockholm University, SE-10691 Stockholm, Sweden
\item[$^{44}$] Dept.~of Physics and Astronomy, Stony Brook University, Stony Brook, NY 11794-3800, USA
\item[$^{45}$] Dept.~of Physics, Sungkyunkwan University, Suwon 440-746, Korea
\item[$^{46}$] Dept.~of Physics and Astronomy, University of Alabama, Tuscaloosa, AL 35487, USA
\item[$^{47}$] Dept.~of Astronomy and Astrophysics, Pennsylvania State University, University Park, PA 16802, USA
\item[$^{48}$] Dept.~of Physics, Pennsylvania State University, University Park, PA 16802, USA
\item[$^{49}$] Dept.~of Physics and Astronomy, Uppsala University, Box 516, S-75120 Uppsala, Sweden
\item[$^{50}$] Dept.~of Physics, University of Wuppertal, D-42119 Wuppertal, Germany
\item[$^{51}$] DESY, D-15738 Zeuthen, Germany
\end{description}

\clearpage

\begin{center}
\par\noindent
{\bf\large The Pierre Auger Collaboration}
\end{center}

\begin{wrapfigure}[9]{l}{0.12\linewidth}
\vspace{-2.9ex}
\includegraphics[width=0.98\linewidth]{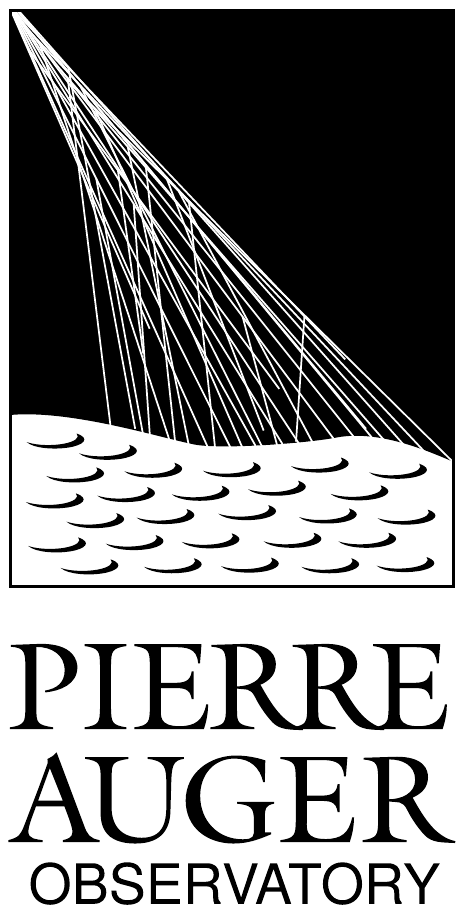}
\end{wrapfigure}
\begin{sloppypar}\noindent

A.~Aab$^{77}$,
P.~Abreu$^{69}$,
M.~Aglietta$^{50,49}$,
I.F.M.~Albuquerque$^{18}$,
I.~Allekotte$^{1}$,
A.~Almela$^{8,11}$,
J.~Alvarez Castillo$^{65}$,
J.~Alvarez-Mu\~niz$^{76}$,
G.A.~Anastasi$^{41,43}$,
L.~Anchordoqui$^{83}$
% ...

\end{sloppypar}

\vspace{1ex}
\begin{center}
\rule{0.1\columnwidth}{0.5pt}
\raisebox{-0.4ex}{\scriptsize$\bullet$}
\rule{0.1\columnwidth}{0.5pt}
\end{center}

\vspace{1ex}

\begin{description}[labelsep=0.2em,align=right,labelwidth=0.7em,labelindent=0em,leftmargin=2em,noitemsep]
\item[$^{1}$] Centro At\'omico Bariloche and Instituto Balseiro (CNEA-UNCuyo-CONICET), San Carlos de Bariloche, Argentina
\item[$^{2}$] Centro de Investigaciones en L\'aseres y Aplicaciones, CITEDEF and CONICET, Villa Martelli, Argentina
\item[$^{3}$] Departamento de F\'\i{}sica and Departamento de Ciencias de la Atm\'osfera y los Oc\'eanos, FCEyN, Universidad de Buenos Aires and CONICET, Buenos Aires, Argentina
\item[$^{4}$] IFLP, Universidad Nacional de La Plata and CONICET, La Plata, Argentina
\item[$^{5}$] Instituto de Astronom\'\i{}a y F\'\i{}sica del Espacio (IAFE, CONICET-UBA), Buenos Aires, Argentina
\item[$^{6}$] Instituto de F\'\i{}sica de Rosario (IFIR) -- CONICET/U.N.R.\ and Facultad de Ciencias Bioqu\'\i{}micas y Farmac\'euticas U.N.R., Rosario, Argentina
\item[$^{7}$] Instituto de Tecnolog\'\i{}as en Detecci\'on y Astropart\'\i{}culas (CNEA, CONICET, UNSAM), and Universidad Tecnol\'ogica Nacional -- Facultad Regional Mendoza (CONICET/CNEA), Mendoza, Argentina
\item[$^{8}$] Instituto de Tecnolog\'\i{}as en Detecci\'on y Astropart\'\i{}culas (CNEA, CONICET, UNSAM), Buenos Aires, Argentina
\item[$^{9}$] Observatorio Pierre Auger, Malarg\"ue, Argentina
\item[$^{10}$] Observatorio Pierre Auger and Comisi\'on Nacional de Energ\'\i{}a At\'omica, Malarg\"ue, Argentina
\item[$^{11}$] Universidad Tecnol\'ogica Nacional -- Facultad Regional Buenos Aires, Buenos Aires, Argentina
\item[$^{12}$] University of Adelaide, Adelaide, S.A., Australia
\item[$^{13}$] Universit\'e Libre de Bruxelles (ULB), Brussels, Belgium
\item[$^{14}$] Centro Brasileiro de Pesquisas Fisicas, Rio de Janeiro, RJ, Brazil
\item[$^{15}$] Centro Federal de Educa\c{c}\~ao Tecnol\'ogica Celso Suckow da Fonseca, Nova Friburgo, Brazil
\item[$^{16}$] Universidade de S\~ao Paulo, Escola de Engenharia de Lorena, Lorena, SP, Brazil
\item[$^{17}$] Universidade de S\~ao Paulo, Instituto de F\'\i{}sica de S\~ao Carlos, S\~ao Carlos, SP, Brazil
\item[$^{18}$] Universidade de S\~ao Paulo, Instituto de F\'\i{}sica, S\~ao Paulo, SP, Brazil
\item[$^{19}$] Universidade Estadual de Campinas, IFGW, Campinas, SP, Brazil
\item[$^{20}$] Universidade Estadual de Feira de Santana, Feira de Santana, Brazil
\item[$^{21}$] Universidade Federal de Pelotas, Pelotas, RS, Brazil
\item[$^{22}$] Universidade Federal do ABC, Santo Andr\'e, SP, Brazil
\item[$^{23}$] Universidade Federal do Paran\'a, Setor Palotina, Palotina, Brazil
\item[$^{24}$] Universidade Federal do Rio de Janeiro, Instituto de F\'\i{}sica, Rio de Janeiro, RJ, Brazil
\item[$^{25}$] Universidade Federal do Rio de Janeiro (UFRJ), Observat\'orio do Valongo, Rio de Janeiro, RJ, Brazil
\item[$^{26}$] Universidade Federal Fluminense, EEIMVR, Volta Redonda, RJ, Brazil
\item[$^{27}$] Universidad Industrial de Santander, Bucaramanga, Colombia
\item[$^{28}$] Charles University, Faculty of Mathematics and Physics, Institute of Particle and Nuclear Physics, Prague, Czech Republic
\item[$^{29}$] Institute of Physics of the Czech Academy of Sciences, Prague, Czech Republic
\item[$^{30}$] Palacky University, RCPTM, Olomouc, Czech Republic
\item[$^{31}$] Institut de Physique Nucl\'eaire d'Orsay (IPNO), Universit\'e Paris-Sud, Univ.\ Paris/Saclay, CNRS-IN2P3, Orsay, France
\item[$^{32}$] Laboratoire de Physique Nucl\'eaire et de Hautes Energies (LPNHE), Universit\'es Paris 6 et Paris 7, CNRS-IN2P3, Paris, France
\item[$^{33}$] Laboratoire de Physique Subatomique et de Cosmologie (LPSC), Universit\'e Grenoble-Alpes, CNRS/IN2P3, Grenoble, France
\item[$^{34}$] Bergische Universit\"at Wuppertal, Department of Physics, Wuppertal, Germany
\item[$^{35}$] Karlsruhe Institute of Technology, Institut f\"ur Experimentelle Kernphysik (IEKP), Karlsruhe, Germany
\item[$^{36}$] Karlsruhe Institute of Technology, Institut f\"ur Kernphysik, Karlsruhe, Germany
\item[$^{37}$] Karlsruhe Institute of Technology, Institut f\"ur Prozessdatenverarbeitung und Elektronik, Karlsruhe, Germany
\item[$^{38}$] RWTH Aachen University, III.\ Physikalisches Institut A, Aachen, Germany
\item[$^{39}$] Universit\"at Hamburg, II.\ Institut f\"ur Theoretische Physik, Hamburg, Germany
\item[$^{40}$] Universit\"at Siegen, Fachbereich 7 Physik -- Experimentelle Teilchenphysik, Siegen, Germany
\item[$^{41}$] Gran Sasso Science Institute (INFN), L'Aquila, Italy
\item[$^{42}$] INAF -- Istituto di Astrofisica Spaziale e Fisica Cosmica di Palermo, Palermo, Italy
\item[$^{43}$] INFN Laboratori Nazionali del Gran Sasso, Assergi (L'Aquila), Italy
\item[$^{44}$] INFN, Sezione di Catania, Catania, Italy
\item[$^{45}$] INFN, Sezione di Lecce, Lecce, Italy
\item[$^{46}$] INFN, Sezione di Milano, Milano, Italy
\item[$^{47}$] INFN, Sezione di Napoli, Napoli, Italy
\item[$^{48}$] INFN, Sezione di Roma "Tor Vergata", Roma, Italy
\item[$^{49}$] INFN, Sezione di Torino, Torino, Italy
\item[$^{50}$] Osservatorio Astrofisico di Torino (INAF), Torino, Italy
\item[$^{51}$] Politecnico di Milano, Dipartimento di Scienze e Tecnologie Aerospaziali , Milano, Italy
\item[$^{52}$] Universit\`a del Salento, Dipartimento di Ingegneria, Lecce, Italy
\item[$^{53}$] Universit\`a del Salento, Dipartimento di Matematica e Fisica ``E.\ De Giorgi'', Lecce, Italy
\item[$^{54}$] Universit\`a dell'Aquila, Dipartimento di Scienze Fisiche e Chimiche, L'Aquila, Italy
\item[$^{55}$] Universit\`a di Catania, Dipartimento di Fisica e Astronomia, Catania, Italy
\item[$^{56}$] Universit\`a di Milano, Dipartimento di Fisica, Milano, Italy
\item[$^{57}$] Universit\`a di Napoli "Federico II", Dipartimento di Fisica ``Ettore Pancini``, Napoli, Italy
\item[$^{58}$] Universit\`a di Roma ``Tor Vergata'', Dipartimento di Fisica, Roma, Italy
\item[$^{59}$] Universit\`a Torino, Dipartimento di Fisica, Torino, Italy
\item[$^{60}$] Benem\'erita Universidad Aut\'onoma de Puebla, Puebla, M\'exico
\item[$^{61}$] Centro de Investigaci\'on y de Estudios Avanzados del IPN (CINVESTAV), M\'exico, D.F., M\'exico
\item[$^{62}$] Unidad Profesional Interdisciplinaria en Ingenier\'\i{}a y Tecnolog\'\i{}as Avanzadas del Instituto Polit\'ecnico Nacional (UPIITA-IPN), M\'exico, D.F., M\'exico
\item[$^{63}$] Universidad Aut\'onoma de Chiapas, Tuxtla Guti\'errez, Chiapas, M\'exico
\item[$^{64}$] Universidad Michoacana de San Nicol\'as de Hidalgo, Morelia, Michoac\'an, M\'exico
\item[$^{65}$] Universidad Nacional Aut\'onoma de M\'exico, M\'exico, D.F., M\'exico
\item[$^{66}$] Institute of Nuclear Physics PAN, Krakow, Poland
\item[$^{67}$] University of \L{}\'od\'z, Faculty of Astrophysics, \L{}\'od\'z, Poland
\item[$^{68}$] University of \L{}\'od\'z, Faculty of High-Energy Astrophysics,\L{}\'od\'z, Poland
\item[$^{69}$] Laborat\'orio de Instrumenta\c{c}\~ao e F\'\i{}sica Experimental de Part\'\i{}culas -- LIP and Instituto Superior T\'ecnico -- IST, Universidade de Lisboa -- UL, Lisboa, Portugal
\item[$^{70}$] ``Horia Hulubei'' National Institute for Physics and Nuclear Engineering, Bucharest-Magurele, Romania
\item[$^{71}$] Institute of Space Science, Bucharest-Magurele, Romania
\item[$^{72}$] University Politehnica of Bucharest, Bucharest, Romania
\item[$^{73}$] Center for Astrophysics and Cosmology (CAC), University of Nova Gorica, Nova Gorica, Slovenia
\item[$^{74}$] Experimental Particle Physics Department, J.\ Stefan Institute, Ljubljana, Slovenia
\item[$^{75}$] Universidad de Granada and C.A.F.P.E., Granada, Spain
\item[$^{76}$] Universidad de Santiago de Compostela, Santiago de Compostela, Spain
\item[$^{77}$] IMAPP, Radboud University Nijmegen, Nijmegen, The Netherlands
\item[$^{78}$] KVI -- Center for Advanced Radiation Technology, University of Groningen, Groningen, The Netherlands
\item[$^{79}$] Nationaal Instituut voor Kernfysica en Hoge Energie Fysica (NIKHEF), Science Park, Amsterdam, The Netherlands
\item[$^{80}$] Stichting Astronomisch Onderzoek in Nederland (ASTRON), Dwingeloo, The Netherlands
\item[$^{81}$] Case Western Reserve University, Cleveland, OH, USA
\item[$^{82}$] Colorado School of Mines, Golden, CO, USA
\item[$^{83}$] Department of Physics and Astronomy, Lehman College, City University of New York, Bronx, NY, USA
\item[$^{84}$] Michigan Technological University, Houghton, MI, USA
\item[$^{85}$] New York University, New York, NY, USA
\item[$^{86}$] Northeastern University, Boston, MA, USA
\item[$^{87}$] Pennsylvania State University, University Park, PA, USA
\item[$^{88}$] University of Chicago, Enrico Fermi Institute, Chicago, IL, USA
\item[$^{89}$] University of Nebraska, Lincoln, NE, USA
\item[] -----
\item[$^{a}$] School of Physics and Astronomy, University of Leeds, Leeds, United Kingdom
\item[$^{b}$] Max-Planck-Institut f\"ur Radioastronomie, Bonn, Germany
\item[$^{c}$] also at Vrije Universiteit Brussels, Brussels, Belgium
\item[$^{d}$] SUBATECH, \'Ecole des Mines de Nantes, CNRS-IN2P3, Universit\'e de Nantes, France
\item[$^{e}$] Fermi National Accelerator Laboratory, USA
\item[$^{f}$] Colorado State University, Fort Collins, CO
\item[$^{g}$] now at Deutsches Elektronen-Synchrotron (DESY), Zeuthen, Germany
\item[$^\dagger$] Deceased
\end{description}

\clearpage

\begin{center}
\par\noindent
{\bf\large The Telescope Array Collaboration}
\end{center}

\begin{wrapfigure}[6]{l}{0.15\linewidth}
\vspace{-2.9ex}
\includegraphics[width=0.98\linewidth]{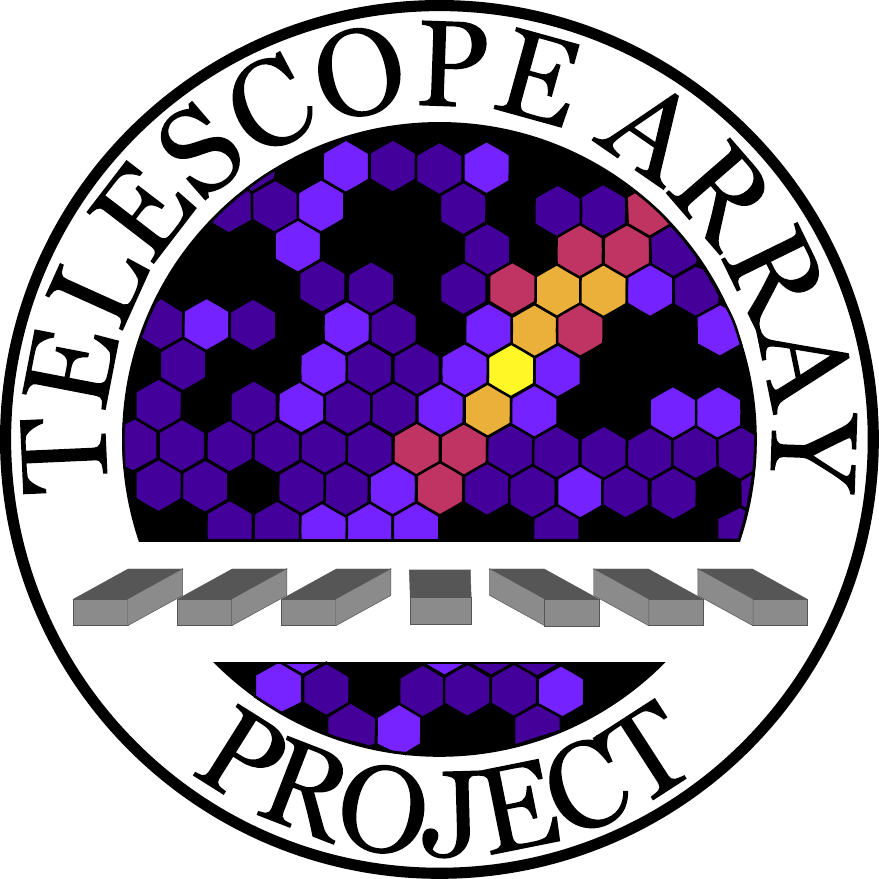}
\end{wrapfigure}
\begin{sloppypar}\noindent
R.U.~Abbasi$^{1}$,
M.~Abe$^{2}$,
T.~Abu-Zayyad$^{1}$,
M.~Allen$^{1}$,
R.~Azuma$^{3}$,
E.~Barcikowski$^{1}$,
J.W.~Belz$^{1}$,
D.R.~Bergman$^{1}$,
S.A.~Blake$^{1}$,
R.~Cady$^{1}$
% ...

\end{sloppypar}

\vspace{1ex}
\begin{center}
\rule{0.1\columnwidth}{0.5pt}
\raisebox{-0.4ex}{\scriptsize$\bullet$}
\rule{0.1\columnwidth}{0.5pt}
\end{center}

\vspace{1ex}
\begin{description}[labelsep=0.2em,align=right,labelwidth=0.7em,labelindent=0em,leftmargin=2em,noitemsep]
\item[$^{1}$] High Energy Astrophysics Institute and Department of Physics and Astronomy, University of Utah, Salt Lake City, Utah, USA
\item[$^{2}$] The Graduate School of Science and Engineering, Saitama University, Saitama, Saitama, Japan
\item[$^{3}$] Graduate School of Science and Engineering, Tokyo Institute of Technology, Meguro, Tokyo, Japan
\item[$^{4}$] Department of Physics and The Research Institute of Natural Science, Hanyang University, Seongdong-gu, Seoul, Korea
\item[$^{5}$] Department of Physics, Tokyo University of Science, Noda, Chiba, Japan
\item[$^{6}$] Department of Physics, Kinki University, Higashi Osaka, Osaka, Japan
\item[$^{7}$] Institute for Cosmic Ray Research, University of Tokyo, Kashiwa, Chiba, Japan
\item[$^{8}$] Kavli Institute for the Physics and Mathematics of the Universe (WPI), Todai Institutes for Advanced Study, the University of Tokyo, Kashiwa, Chiba, Japan
\item[$^{9}$] Graduate School of Science, Osaka City University, Osaka, Osaka, Japan
\item[$^{10}$] Information Engineering Graduate School of Science and Technology, Shinshu University, Nagano, Nagano, Japan
\item[$^{11}$] Faculty of Engineering, Kanagawa University, Yokohama, Kanagawa, Japan
\item[$^{12}$] Interdisciplinary Graduate School of Medicine and Engineering, University of Yamanashi, Kofu, Yamanashi, Japan
\item[$^{13}$] Astrophysical Big Bang Laboratory, RIKEN, Wako, Saitama, Japan
\item[$^{14}$] Department of Physics, Tokyo City University, Setagaya-ku, Tokyo, Japan
\item[$^{15}$] Institute for Nuclear Research of the Russian Academy of Sciences, Moscow, Russia
\item[$^{16}$] Advanced Research Institute for Science and Engineering, Waseda University, Shinjuku-ku, Tokyo, Japan
\item[$^{17}$] Department of Physics, Chiba University, Chiba, Chiba, Japan
\item[$^{18}$] Department of Physics, School of Natural Sciences, Ulsan National Institute of Science and Technology, UNIST-gil, Ulsan, Korea
\item[$^{19}$] Department of Physics, Yonsei University, Seodaemun-gu, Seoul, Korea
\item[$^{20}$] Institute of Particle and Nuclear Studies, KEK, Tsukuba, Ibaraki, Japan
\item[$^{21}$] Faculty of Science, Kochi University, Kochi, Kochi, Japan
\item[$^{22}$] Department of Physical Sciences, Ritsumeikan University, Kusatsu, Shiga, Japan
\item[$^{23}$] Department of Physics, Sungkyunkwan University, Jang-an-gu, Suwon, Korea
\item[$^{24}$] Sternberg Astronomical Institute, Moscow M.V.~Lomonosov State University, Moscow, Russia
\item[$^{25}$] Academic Assembly School of Science and Technology Institute of Engineering, Shinshu University, Nagano, Nagano, Japan
\item[$^{26}$] Department of Physics and Astronomy, Rutgers University - The State University of New Jersey, Piscataway, New Jersey, USA
\item[$^{27}$] Earthquake Research Institute, University of Tokyo, Bunkyo-ku, Tokyo, Japan
\item[$^{28}$] Graduate School of Information Sciences, Hiroshima City University, Hiroshima, Hiroshima, Japan
\item[$^{29}$] Service de Physique Th$\acute{\rm e}$orique, Universit$\acute{\rm e}$ Libre de Bruxelles, Brussels, Belgium
\item[$^{30}$] National Institute of Radiological Science, Chiba, Chiba, Japan
\item[$^{31}$] National Institute of Chemical Physics and Biophysics, Estonia
\item[$^{32}$] Department of Physics and Institute for the Early Universe, Ewha Womans University, Seodaaemun-gu, Seoul, Korea
\item[$^{33}$] Department of Physics, Ehime University, Matsuyama, Ehime, Japan
\item[] -----
\item[$^\dagger$] Deceased
\end{description}

\clearpage

\section*{Acknowledgments of the IceCube Collaboration}

\begin{sloppypar}
The authors gratefully acknowledge the support from the following agencies
and institutions: USA -- U.S.\ National Science Foundation-Office of Polar Programs, U.S.\ National
Science Foundation-Physics Division, University of Wisconsin Alumni Research Foundation, the
Center for High Throughput Computing (CHTC) at the University of Wisconsin -- Madison, the
Open Science Grid (OSG) grid infrastructure and the Extreme Science and Engineering Discovery
Environment (XSEDE); U.S.\ Department of Energy, and National Energy Research Scientific
Computing Center; Particle Astrophysics research computing center at the University of Maryland;
Institute for Cyber-Enabled Research at Michigan State University; Astroparticle Physics Computational
Facility at Marquette University; Belgium -- Funds for Scientific Research (FRS-FNRS and
FWO), FWO Odysseus and Big Science programs, Belgian Federal Science Policy Office (Belspo);
Germany -- Bundesministerium für Bildung und Forschung (BMBF), Deutsche Forschungsgemeinschaft
(DFG), Helmholtz Alliance for Astroparticle Physics (HAP), Initiative and Networking Fund
of the Helmholtz Association; Deutsches Elektronen Synchrotron (DESY); Cluster of Excellence
(PRISMA ? EXC 1098); High Performance Computing Cluster of the IT-Center of the RWTH
Aachen; Sweden -- Swedish Research Council, Swedish Polar Research Secretariat, Swedish National
Infrastructure for Computing (SNIC), and Knut and Alice Wallenberg Foundation; Canada --
Natural Sciences and Engineering Research Council of Canada, Calcul Québec, Compute Ontario,
WestGrid and Compute Canada; Denmark -- Villum Fonden, Danish National Research Foundation
(DNRF); New Zealand -- Marsden Fund, New Zealand; Australian Research Council; Japan -- Japan
Society for Promotion of Science (JSPS) and Institute for Global Prominent Research (IGPR) of
Chiba University; Korea -- National Research Foundation of Korea (NRF); Switzerland -- Swiss
National Science Foundation (SNSF).
\end{sloppypar}

\section*{Acknowledgments of the Pierre Auger Collaboration}


\begin{sloppypar}
The successful installation, commissioning, and operation of the Pierre
Auger Observatory would not have been possible without the strong
commitment and effort from the technical and administrative staff in
Malarg\"ue. We are very grateful to the following agencies and
organizations for financial support:
\end{sloppypar}

\begin{sloppypar}
Argentina -- Comisi\'on Nacional de Energ\'\i{}a At\'omica; Agencia Nacional de
Promoci\'on Cient\'\i{}fica y Tecnol\'ogica (ANPCyT); Consejo Nacional de
Investigaciones Cient\'\i{}ficas y T\'ecnicas (CONICET); Gobierno de la
Provincia de Mendoza; Municipalidad de Malarg\"ue; NDM Holdings and Valle
Las Le\~nas; in gratitude for their continuing cooperation over land
access; Australia -- the Australian Research Council; Brazil -- Conselho
Nacional de Desenvolvimento Cient\'\i{}fico e Tecnol\'ogico (CNPq);
Financiadora de Estudos e Projetos (FINEP); Funda\c{c}\~ao de Amparo \`a
Pesquisa do Estado de Rio de Janeiro (FAPERJ); S\~ao Paulo Research
Foundation (FAPESP) Grants No.~2010/07359-6 and No.~1999/05404-3;
Minist\'erio de Ci\^encia e Tecnologia (MCT); Czech Republic -- Grant
No.~MSMT CR LG15014, LO1305, LM2015038 and
CZ.02.1.01/0.0/0.0/16\_013/0001402; France -- Centre de Calcul
IN2P3/CNRS; Centre National de la Recherche Scientifique (CNRS); Conseil
R\'egional Ile-de-France; D\'epartement Physique Nucl\'eaire et Corpusculaire
(PNC-IN2P3/CNRS); D\'epartement Sciences de l'Univers (SDU-INSU/CNRS);
Institut Lagrange de Paris (ILP) Grant No.~LABEX ANR-10-LABX-63 within
the Investissements d'Avenir Programme Grant No.~ANR-11-IDEX-0004-02;
Germany -- Bundesministerium f\"ur Bildung und Forschung (BMBF); Deutsche
Forschungsgemeinschaft (DFG); Finanzministerium Baden-W\"urttemberg;
Helmholtz Alliance for Astroparticle Physics (HAP);
Helmholtz-Gemeinschaft Deutscher Forschungszentren (HGF); Ministerium
f\"ur Innovation, Wissenschaft und Forschung des Landes
Nordrhein-Westfalen; Ministerium f\"ur Wissenschaft, Forschung und Kunst
des Landes Baden-W\"urttemberg; Italy -- Istituto Nazionale di Fisica
Nucleare (INFN); Istituto Nazionale di Astrofisica (INAF); Ministero
dell'Istruzione, dell'Universit\'a e della Ricerca (MIUR); CETEMPS Center
of Excellence; Ministero degli Affari Esteri (MAE); Mexico -- Consejo
Nacional de Ciencia y Tecnolog\'\i{}a (CONACYT) No.~167733; Universidad
Nacional Aut\'onoma de M\'exico (UNAM); PAPIIT DGAPA-UNAM; The Netherlands
-- Ministerie van Onderwijs, Cultuur en Wetenschap; Nederlandse
Organisatie voor Wetenschappelijk Onderzoek (NWO); Stichting voor
Fundamenteel Onderzoek der Materie (FOM); Poland -- National Centre for
Research and Development, Grants No.~ERA-NET-ASPERA/01/11 and
No.~ERA-NET-ASPERA/02/11; National Science Centre, Grants
No.~2013/08/M/ST9/00322, No.~2013/08/M/ST9/00728 and No.~HARMONIA
5--2013/10/M/ST9/00062, UMO-2016/22/M/ST9/00198; Portugal -- Portuguese
national funds and FEDER funds within Programa Operacional Factores de
Competitividade through Funda\c{c}\~ao para a Ci\^encia e a Tecnologia
(COMPETE); Romania -- Romanian Authority for Scientific Research ANCS;
CNDI-UEFISCDI partnership projects Grants No.~20/2012 and No.~194/2012
and PN 16 42 01 02; Slovenia -- Slovenian Research Agency; Spain --
Comunidad de Madrid; Fondo Europeo de Desarrollo Regional (FEDER) funds;
Ministerio de Econom\'\i{}a y Competitividad; Xunta de Galicia; European
Community 7th Framework Program Grant No.~FP7-PEOPLE-2012-IEF-328826;
USA -- Department of Energy, Contracts No.~DE-AC02-07CH11359,
No.~DE-FR02-04ER41300, No.~DE-FG02-99ER41107 and No.~DE-SC0011689;
National Science Foundation, Grant No.~0450696; The Grainger Foundation;
Marie Curie-IRSES/EPLANET; European Particle Physics Latin American
Network; European Union 7th Framework Program, Grant
No.~PIRSES-2009-GA-246806; European Union's Horizon 2020 research and
innovation programme (Grant No.~646623); and UNESCO.
\end{sloppypar}

\section*{Acknowledgments of the Telescope Array Collaboration}

\begin{sloppypar}
The Telescope Array experiment is supported by the Japan Society for the
Promotion of Science through Grants-in-Aid for Scientific Research on Specially
Promoted Research (21000002) ``Extreme Phenomena in the Universe Explored by
Highest Energy Cosmic Rays'' and for Scientific Research (19104006), and the
Inter-University Research Program of the Institute for Cosmic Ray Research; by
the U.S.\ National Science Foundation awards PHY-0601915, PHY-1404495,
PHY-1404502, and PHY1607727; by the National Research Foundation of Korea
(2015\-R1A2A1A\-0100\-6870, 2015R1A2A1A15055344, 2016R1A5A1013277, 2007-0093860,
2016R1A2B4014967); by the Russian Academy of Sciences, RFBR grant 16-02-00962a
(INR), IISN project No.\ 4.4502.13, and Belgian Science Policy under IUAP
VII/37 (ULB). The foundations of Dr.\ Ezekiel R.\ and Edna Wattis Dumke,
Willard L.\ Eccles, and George S.\ and Dolores Dor\'e Eccles all helped with
generous donations. The State of Utah supported the project through its
Economic Development Board, and the University of Utah through the Office of
the Vice President for Research. The experimental site became available through
the cooperation of the Utah School and Institutional Trust Lands Administration
(SITLA), U.S.\ Bureau of Land Management (BLM), and the U.S.\ Air Force. We
appreciate the assistance of the State of Utah and Fillmore offices of the BLM
in crafting the Plan of Development for the site. Patrick Shea assisted the
collaboration with valuable advice on a variety of topics. The people and the
officials of Millard County, Utah have been a source of steadfast and warm
support for our work which we greatly appreciate.  We are indebted to the
Millard County Road Department for their efforts to maintain and clear the
roads which get us to our sites. We gratefully acknowledge the contribution
from the technical staffs of our home institutions. An allocation of computer
time from the Center for High Performance Computing at the University of Utah
is gratefully acknowledged.
\end{sloppypar}

\newpage


\clearpage

\addPaper{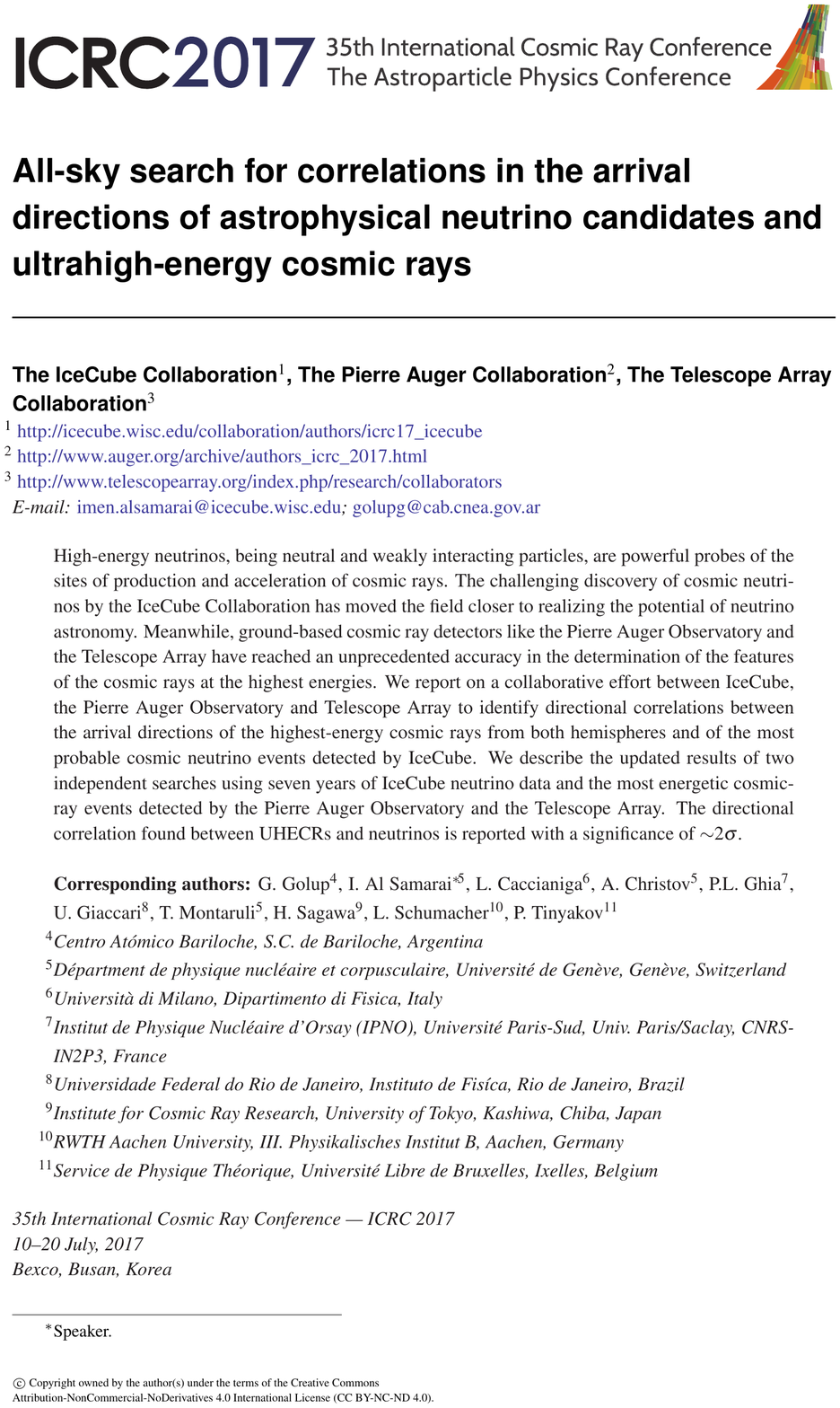}{}{All-sky search for correlations in the arrival directions of astrophysical neutrino candidates and ultrahigh-energy cosmic rays}

\end{document}